\documentclass[aip,jcp,reprint,groupedaddress,twocolumn]{revtex4-1}
\usepackage{graphicx}
\usepackage{epsfig}
\usepackage{amsmath}
\newcommand{\bra}[1]{\ensuremath{\langle{#1}|\,}}
\newcommand{\ket}[1]{\ensuremath{\,|{#1}\rangle}}
\newcommand{\braket}[1]{\ensuremath{\langle{#1}\rangle}}

\begin{document}

\title{Second-order post-Hartree-Fock perturbation theory for the electron current }

\author{Alan A. Dzhioev}
\altaffiliation{On leave of absence from Bogoliubov Laboratory of Theoretical Physics, Joint Institute for Nuclear Research,  RU-141980 Dubna, Russia }
\author{D. S. Kosov}
\email{dkosov@ulb.ac.be}
\affiliation{Department of Physics, 
Universit\'e Libre de Bruxelles, Campus Plaine, CP 231, Blvd du Triomphe, B-1050 Brussels, Belgium }


\begin{abstract}
Based on the super-fermion representation  of quantum kinetic equations we develop nonequilibrium, post-Hartree-Fock many-body perturbation theory for the current through a region of interacting electrons. We apply the theory to out of equilibrium Anderson model and
discuss practical implementation of the approach. Our calculations show that nonequilibrium electronic correlations  may produce significant quantitative and qualitative corrections to mean-field electronic transport properties. We find that the nonequilibrium leads to  enhancement of electronic correlations.
\end{abstract}

\maketitle

\section{Introduction}
The second order M$\o$ller-Plesset perturbation theory (MP2) is a very powerful tool for practical quantum chemical calculation.\cite{jensen-book} MP2 accounts  in many cases for the large fraction of post-Hartree-Fock electronic correlations. It
gives quite accurate results for intermolecular interactions, molecular geometries, dissociation energies.\cite{jensen-book}  In this paper, we propose the extension of the MP2 theory to nonequilibrium, namely, we develop the second order post-Hartree-Fock perturbation theory for the electron current through a  region of out of equilibrium, interacting electrons. As an example, we consider electron transport through  Anderson model, but all our derivations are also directly applicable and transferable to a general  molecular system connected to two metal electrodes and  described by full many-electron Hamiltonian.

Recently, there have been several attempts to include post-Hartree-Fock electronic correlation into  transport calculations through molecules.\cite{PhysRevB.77.115333,PhysRevB.80.115107,PhysRevB.79.155110,PhysRevB.80.165305,thygesen07,PhysRevB.75.075102} These approaches are mostly based on various implementation of nonequilibrium Green's function based GW theory.  Our method is quite different. We start with kinetic equations for many-body density matrix. We convert the kinetic equation to the super-Fock space, separate Hartree-Fock and the normal ordered parts of the many-particle Liouvillian, and  then develop nonequilibrium many body perturbation theory much along the lines of the traditional MP2 theory.

The rest of the paper is organized as follows. In Section II, we describe quantum kinetic equation and super-fermion formalism. Section III presents
the main equations of nonequilibrium second-order many-body perturbation theory. Section IV describes numerical calculations.
Conclusions are given in Section V. We use natural units throughout the paper: $\hbar= k_B = |e| = 1$, where $-|e|$ is the electron charge.

\section{Lindblad kinetic equation and super-Fock space}
To be specific in our discussion, let us consider one spin-degenerate level (impurity) attached to two macroscopic leads.
We begin with the Lindblad kinetic equation
\begin{align}
\label{lindblad}
&  \frac{d\rho(t)}{dt} =-i[H,\rho(t)]
  \nonumber
  \\
 &  +{\sum_{\sigma k \alpha}}  \sum_{ \mu=1,2} (2L_{\sigma k \alpha \mu} \rho(t) L^\dag_{\sigma k \alpha \mu}
   -\{ L^\dag_{\sigma k \alpha \mu} L_{\sigma k \alpha \mu},\rho(t) \}).
\end{align}
Here $H$ is the { \it finite} Hamiltonian for the embedded impurity  (the "embedded" here means that the
systems consists of the impurity itself and  finite parts of left/right leads).
The Hamiltonian for the embedded impurity  can be written in the following form:
\begin{align}\label{embedded}
  H&= H_{S} + H_{B} + H_{T},
  \notag\\
  H_{S}&= \varepsilon_0 \sum_{\sigma} n_\sigma + U n_\uparrow n_\downarrow,
 \notag\\
 H_B&=\sum_{\sigma k \alpha} \varepsilon_{k\alpha} a^{\dagger}_{\sigma k \alpha} a_{\sigma k \alpha},
 \notag\\
  H_{T}&= -t \sum_{\sigma k \alpha}( a^{\dagger}_{\sigma k \alpha} a_\sigma + h.c.),
\end{align}
where $a^\dag$, $a$ are fermionic creation and annihilation operators: $a^\dag_\sigma$ ($a_\sigma$) creates (destroys) an electron with spin $\sigma$ in the impurity and
$a^\dag_{\sigma k \alpha}$ ($a_{\sigma k \alpha}$) creates (destroys) an electron of energy $\varepsilon_{k}$ and spin $\sigma=\uparrow, \downarrow$ in the left ($\alpha=L$) or right ($\alpha=R$) lead;
$n_\sigma$ is the number operator for electrons of spin $\sigma$ in the impurity; $U$ is the Coulomb  interaction between two electrons of opposite spin in the impurity;
 $t$ is the tunneling matrix element between the states in the leads and in the impurity. For simplicity we assume that the tunneling matrix element
is real and independent of $\alpha$, $k$ and $\sigma$. We also assume that the left and right leads are identical.
Here $\sum\limits_{k\alpha}$ runs over $2N$
 {\it discrete} single particle levels~$k \alpha $ coupled  directly to the Lindblad dissipators.

The  dissipators have the following form:
\begin{align}
  L_{\sigma k\alpha1}=\sqrt{\Gamma_{k\alpha1}} a_{\sigma k\alpha},~~L_{\sigma k\alpha2}=\sqrt{\Gamma_{k\alpha2}} a^\dag_{\sigma k\alpha},
\end{align}
with $\Gamma_{k\alpha1}=\gamma_{k\alpha} (1- f_{k\alpha})$, $\Gamma_{k\alpha2}=\gamma_{k\alpha} f_{k\alpha}$.
Here
$f_{k\alpha}=[1+\exp[(\varepsilon_{k\alpha}-\mu_\alpha)/T_{\alpha}]^{-1}$ and $\gamma_{k\alpha}$  is given by the imaginary part of the electrode self energy
$\gamma_{k\alpha} = - \frac{1}{2} \text{Im}[\Sigma_{k \alpha}] $.\cite{gurvitz96,PhysRevB.74.235309}
The Lamb shift is included into the single-particle energt of the leads:
$ \varepsilon_{k \alpha}  = \varepsilon_{k \alpha}^{(0)} + \text{Re}[\Sigma_{k\alpha} ]$, where  $\varepsilon_{k\alpha}^{(0)}$ is bare energy of single-particle levels in the leads.

We would like to emphasize here that for the  steady state electron transport calculations the Lindblad master equation can be made as accurate  and  exact as necessary by the increasing the number of lead states $k$ included into the Hamiltonian for the embedded impurity.
So the master equation (\ref{lindblad}) can be considered as the {\it exact} starting point for first principles electronic transport calculations.

Let us begin to work with the kinetic equation (\ref{lindblad}). First, we re-write it in the super-fermion representation.
For many-particle quantum systems the density matrix $\rho(t)$ and the Hamiltonian are operators in the Fock space. The Fock space can be defined by some orthonormal complete set of basis vectors:
\begin{equation}
\sum_n |n \rangle \langle n| =I,\;  \;\;  \langle n| m \rangle = \delta_{nm}.
\label{fock}
\end{equation}
Let us introduce the additional Fock space which is identical copy of the initial Fock space
\begin{equation}
\sum_n |\widetilde{n} \rangle \langle \widetilde{n}| =\widetilde{I},\;  \;\;  \langle \widetilde{n}| \widetilde{m} \rangle = \delta_{nm}.
\label{tilde-fock}
\end{equation}
We denote all vectors and operators in this additional Fock space by "tilde".
The vectors $|n\rangle$ and $|\widetilde{n}\rangle $ span the so-called super-Fock space, which is a direct product of the original and the "tilde" Fock spaces.
In the super-Fock space we define  "left vacuum vector"
\begin{equation}
|I \rangle = \sum_{n} \alpha_n|n \rangle \otimes  |\widetilde{n}\rangle,
\label{unit-vector}
\end{equation}
and  "nonequilibrium wavefunction":
\begin{equation}
|\rho(t)\rangle =\rho(t) | I \rangle,
\label{vacuum}
\end{equation}
where $\alpha_n$ ($|\alpha_n|=1$) is an arbitrary phase factor.
Now the nonequilibrium average can be written as
 \begin{equation}\label{expect}
 \langle A(t) \rangle = \mathrm{Tr} [A \rho(t) ] = \langle I | A  |\rho(t)\rangle .
\end{equation}

We consider a system which consists of  fermions distributed over $N$ levels. If we take vector $|n\rangle$ and $|\widetilde{n}\rangle$  to be the particle number eigenstate,
$|n \rangle = |n_1 n_2\ldots n_N\rangle$, $|\widetilde{n} \rangle = \widetilde{ |n_1 n_2\ldots n_N\rangle}$,
and choose $\alpha_n=(-i)^{n_1+n_2 + \ldots+n_N}$, then
we can readily demonstrate by the straightforward algebraic manipulations that
\begin{equation}
a_j |I\rangle = -i \widetilde{a}^{\dagger}_j |I\rangle, \;\;\;\;\; a^{\dagger}_j |I\rangle = -i \widetilde{a}_j |I\rangle.
\label{tilde}
\end{equation}
This is so-called  "tilde conjugation rule".  It
transforms the original operators to the tilde operators and it is one of the most important relations. In particular,
it follows from~\eqref{tilde} that the left vacuum vector, $\bra{I}$, is always the vacuum for $a^\dag_j-i\widetilde{a}_j$  and $\widetilde a^\dag_j+i a_j$ operators.
Moreover, it shows that, in some sense,
the particle creation (annihilation) in ordinary Fock space is equivalent to the particle  annihilation (creation) in the "tilde" space.

We can formulate the following rules for the operations in the super-Fock space:\cite{schmutz78,prosen08,Harbola2008,dzhioev11a}
\begin{enumerate}
\item
The  left vacuum vector $| I \rangle $ and the nonequilibrium wavefunction $ |\rho(t)\rangle =\rho(t) | I \rangle $  are invariant under the tilde conjugation
$\widetilde{|\rho(t) \rangle }  = |\rho(t) \rangle $,  $\widetilde{| I \rangle }  = | I \rangle $, and $\bra{I}\rho(t)\rangle=1$.

\item Tilde conjugation rules: The main rule is
$
a_j |I\rangle = -i \widetilde{a}^{\dagger} |I\rangle, \;\;\; a^{\dagger}_j |I\rangle = -i \widetilde{a} |I\rangle
$
and as consequence the double tilde conjugation does not change the operator $\widetilde{\widetilde{A}}  =A$
and
$
\widetilde{(c_1 A + c_2 B)} = c_1^* \widetilde{A} + c^*_2 \widetilde{B}
$.

\item Evolution of the system is described by the time-dependent Schr\"odinger equation
$
i\frac{d}{dt} |\rho(t)\rangle = L |\rho(t)\rangle,
$
where the Liouvillian is obtained from the  corresponding master equation for the density matrix by the  tilde conjugation rules.
The nonequilibrium average is given by $\langle A(t) \rangle = \langle I | A |\rho(t) \rangle $ and $ \langle I | L =0$.

\end{enumerate}

If we act by  the Lindblad equation on the "left vacuum vector" $|I\rangle$, employ the tilde conjugation rules  and use the fact that the density matrix
$\rho= \rho(a^{\dagger}, a) $ is the operator in original Fock space therefore it commutes with all "tilde" operators,
then the Lindblad master equation becomes the time-dependent Schr\"odinger equation in the super-Fock space
\begin{equation}
i \frac{d}{dt} |\rho(t) \rangle = L |\rho(t) \rangle
\end{equation}
where
the Liouvillian $L$ is given by
\begin{align}\label{H_lind}
 L& =L_S +  L_B +  L_T,
\end{align}
where $L_S = H_S-\widetilde H_S$ and $L_T = H_T-\widetilde H_T$ are both Hermitian, while
\begin{equation}
L_B = H_B - \widetilde H_B-i \sum_{\sigma k\alpha}\Pi_{\sigma k\alpha}
\end{equation}
includes the non-Hermitian part which is responsible for the dissipation in the system:
\begin{align} \label{Pi}
\Pi_{\sigma k\alpha} =&(\Gamma_{k\alpha1}-\Gamma_{k\alpha2})( a^\dag_{\sigma k\alpha}a_{\sigma k\alpha}+
  \widetilde a^\dag_{\sigma k\alpha}\widetilde a_{\sigma k\alpha})
  \notag\\&-
  2i(\Gamma_{k\alpha1}\widetilde a_{\sigma k\alpha}a_{\sigma k\alpha}+\Gamma_{k\alpha2}\widetilde a^\dag_{\sigma k\alpha}a^\dag_{\sigma k\alpha})+2\Gamma_{k\alpha2}.
   \end{align}
In the above equations all tilde operators are obtained from non-tilde ones by the tilde conjugation rules.

In this paper, we  focus on nonequilibrium steady state,
where the density matrix  $|\rho(t)\rangle $  has already reached its asymptotic steady state  $|\rho_{\infty} \rangle $ and  does not anymore depend on time. Therefore, the electron transport problem is reduced to the problem of  finding the eigenvector
 with zero eigenvalue of complex, non-Hermitian, finite-dimensional Liouville operator~\eqref{H_lind} acting in the super-Fock space
\begin{equation}
L |\rho_\infty\rangle =0.
\label{lrho=0}
\end{equation}
Once the steady state density matrix is found, we can compute the current as
\begin{equation}\label{current}
J_\alpha = -i t\sum_{\sigma k}\bra{I}( a^\dag_{\sigma k\alpha} a_\sigma -  a^\dag_\sigma a_{\sigma k\alpha}) \ket{\rho_\infty}.
\end{equation}
In the next section, we show how one can find $ |\rho_\infty\rangle$ perturbatively beyond nonequilibrium Hartree-Fock approximation.

\section{Nonequilibrium many-body perturbation theory}

Before we start to develop many-body perturbation theory, we make the important remark on the notation abuse  in this section of the paper: only creation/annihilation operators written with letter $a$ (such as for example $a_{k \alpha}$ and $a^{\dagger}_{k \alpha}$) are related to each other by the hermitian conjugation; all other creation  $c^{\dagger}$ and annihilation $c$ operators are "canonically conjugated" to each other, i.e., for example,  $c^{\dagger}$ does not mean $(c)^{\dagger}$ although $\{ c,c^{\dagger}\}=1$.

Let us first perform the normal ordering of the Liouvillian, and
find the nonequilibrium Hartree-Fock density matrix $\ket{\rho_\infty^{(0)}}$ and
corresponding nonequilibrium quasiparticles (i.e. quasiparticles, which have the Hartree-Fock density matrix as a vacuum in the super-Fock space).

Using the Wick theorem we  rewrite the Liouvillian~\eqref{H_lind}  as
\begin{equation}
  L =  L^{(0)} + L',
\end{equation}
where $L'$ is the normal ordered part of the Liouvillian
\begin{equation}
L'= U:(a^\dag_\uparrow a_\uparrow a^\dag_\downarrow a_\downarrow - \mathrm{t.c.}):,
 \label{:l:}
\end{equation}
and the notation (t.c.) means the tilde conjugation (i.e. $a_\sigma \to \widetilde a_\sigma$, etc.).  The normal
ordering is asymmetric: it is performed with respect to $\bra{I}$ from the left and the nonequilibrium Hartree-Fock density matrix $\ket{\rho_\infty^{(0)}}$ from the right.\cite{dzhioev11a}

The mean-field part of $L$ is
\begin{equation}\label{mean_field}
L^{(0)}=L^{(0)}_S+L_T+ L_B,
\end{equation}
where $L^{(0)}_S$ is a quadratic part of  $L_S$  in the  Hartree-Fock approximation
\begin{equation}
  L^{(0)}_S =  \sum_\sigma( \varepsilon_0 + U\langle n_{-\sigma}\rangle)(a^\dag_\sigma a_\sigma-\widetilde a^\dag_\sigma \widetilde a_\sigma).
\end{equation}
The mean-field population $\langle n_\sigma\rangle=\bra{I}n_\sigma\ket{\rho_\infty^{(0)}}$ is spin independent  and $\braket{a^\dag_\sigma a_{-\sigma}}=0$ due to the symmetry of the problem.
The quadratic part of the Liouvillian, $L^{(0)}$,
can be diagonalized exactly
in terms of nonequilibrium quasiparticle creation and annihilation operators.\cite{dzhioev11a} As a result we get
\begin{equation}
  L^{(0)} = \sum_{n \sigma}  (\Omega_n c^\dag_{\sigma n}c_{\sigma n} - \Omega^*_n \widetilde c^\dag_{\sigma n}\widetilde c_{\sigma n}),
\end{equation}
where the nonequilibrium quasiparticles are defined as
\begin{align}\label{a_to_c}
 c^\dag_{\sigma n} &= \psi_{n}  ( a^\dag_\sigma - i \widetilde a_{\sigma}) +\sum_{k\alpha}\psi_{n,k\alpha}  ( a^\dag_{\sigma k\alpha}  - i \widetilde a_{k\alpha}),
  \notag\\
 c_{\sigma n} &=\psi_{n} a_\sigma +  i  \varphi_{n} ( \widetilde a^\dag_\sigma + i  a_{\sigma})
  \notag\\
& +\sum_{k\alpha}(\psi_{n,k\alpha}  a_{\sigma k\alpha} +i \varphi_{n,k\alpha}
 ( \widetilde a^\dag_{\sigma k\alpha}  + i a_{k\alpha})
 ),
\end{align}
and $\widetilde c^\dag_{\sigma n}, \widetilde c_{\sigma n} $ are obtained from $c^\dag_{\sigma n}, c_{\sigma n} $ by the tilde conjugation rules.
The operators $c^\dag,~c,~\widetilde c^\dag,~\widetilde c$ are connected to $a^\dag,~a,~\widetilde a^\dag,~\widetilde a$ by
canonical (but not unitary) transformations. Nonequilibrium quasiparticle creation and annihilation operators
obey the fermionic anticommutation relations.
By the construction,  $\langle I|$ is the left  vacuum for $c^\dag_{n},~\widetilde c^\dag_{n}$ operators.  The vacuum state for  $c_{n},~\widetilde c_{n}$ operators, $\ket{\rho^{(0)}_\infty}$, is automatically
the  zero-eigenvalue eigenstate of the mean-field Liouvillian $L^{(0)}$, and therefore, it is
the steady state density matrix in the Hartree-Fock approximation:
\begin{equation}
  L^{(0)}\ket{\rho_\infty^{(0)}}=0,~~~\bra{I}\rho_\infty^{(0)}\rangle=1.
\end{equation}
In other words, the mean-field  density matrix is the density matrix which does not contain nonequilibrium quasiparticle excitations.

The  amplitudes $\psi_{n}$, $\psi_{n,k\alpha}$ and quasiparticle energies $\Omega_n$ ($n=1,\ldots2N+1$) are the solution of the following eigenvalue
problem (obtained from the equations-of-motion $[L,c^{\dagger}_{\sigma n} ] = \Omega_n c^{\dagger}_{\sigma n}$)
\begin{align}\label{sys3}
  &\varepsilon_{HF} \psi_{n}-t \sum\limits_{k\alpha}\psi_{n,k\alpha}=\Omega_n \psi_{n},
  \notag\\
  &E_{k\alpha}\psi_{n,k\alpha}-  t\psi_{n}=\Omega_n \psi_{n,k\alpha},
\end{align}
where $E_{k\alpha}=\varepsilon_{k\alpha}-i\gamma_{k\alpha}$ and $\varepsilon_{HF}=\varepsilon_0 + U\langle n_{\sigma}\rangle$  is the Hartree-Fock single particle energy.
The  amplitudes $\psi_{n}$, $\psi_{n,k\alpha}$ are normalized as
\begin{equation}
  \psi_{n}\psi_{n'}+\sum_{k\alpha} \psi_{n,\,k\alpha}\psi_{{n'},\,k\alpha}=\delta_{nn'}.
\end{equation}
The  amplitudes $\varphi_n$, $\varphi_{n,k \alpha}$ satisfy the following nonhomogeneous  system of linear equations (obtained from  $[L,c_{\sigma n} ] = -\Omega_n c_{\sigma n}$):
\begin{align}\label{sys4}
   &(\varepsilon_{HF} - \Omega_n) \varphi_{n}-t \sum_{k\alpha} \varphi_{n,k\alpha}=t \sum_{k\alpha} f_{k\alpha} \psi_{n,k\alpha},
   \notag\\
  &(E^*_{k\alpha}-\Omega_n)\varphi_{n,k\alpha}-  t\varphi_{n}=-t f_{k\alpha} \psi_{n}.
\end{align}
This is  a set of nonlinear equations since the Hartree-Fock single particle energy $\varepsilon_{HF}$ depends on amplitudes\begin{equation}\label{population}
\langle n_\sigma \rangle=\sum_n \psi_n\varphi_n,
\end{equation}
which in turn depends on $\varepsilon_{HF}$. We solve these equations  numerically by self-consistent iterations.  We first guess the population
$\langle n_\sigma \rangle$, then for a given $\varepsilon_{HF}$ we solve the eigenvalue problem~\eqref{sys3}. Then we solve the linear system of equations~\eqref{sys4} with the known quasiparticle spectrum $\Omega_n$  and known amplitudes  $\psi_{n}$, $\psi_{n,k\alpha}$. Then we compute new nonequilibrium mean-field population $\langle n_\sigma \rangle$ of the impurity and continue this loop for self-consistent iteration until the variation in the impurity population between the iterations becomes negligible.

Now we are in a position to develop the post-Hartree-Fock many-body perturbation theory, which takes into account the normal ordered part of the Liouvillian  (\ref{:l:}).
We introduce the continuous real parameter $\lambda$, which
will be set to unity in the end of the calculations
\begin{equation}
  L =  L^{(0)} + \lambda L'.
\end{equation}
The normal  ordered post-Hartree-Fock perturbation $L'$ in the nonequilibrium quasiparticle basis is given in the Appendix.
 We expand  the exact steady state density matrix in powers of $\lambda$,
\begin{equation}\label{ss_expansion}
  \ket{\rho_\infty} = \ket{\rho^{(0)}_\infty} + \lambda \ket{\rho^{(1)}_\infty} +  \lambda^2 \ket{\rho^{(2)}_\infty}+\ldots,
\end{equation}
where $\bra{I}\rho_\infty^{(p)}\rangle=0,~p\ge1$. If we require that $ \ket{\rho_\infty} $ is the steady state density matrix for the full Liouvillian  (\ref{lrho=0}), we get
the following recurrent relations for the  $p$th-order correction to the Hartree-Fock density matrix:
\begin{equation}\label{pert_eq}
 L_{0} \ket{\rho^{(p)}_\infty} = -  L'  \ket{\rho^{(p-1)}_\infty}.
\end{equation}
In the case of the second-order perturbation theory this recurrent relation simply becomes the systems of two equations
\begin{eqnarray}
L_0 \ket{\rho^{(1)}_\infty} &&=- L' \ket{\rho^{(0)}_\infty},
\label{mp2-1}
\\
 L_0  \ket{\rho^{(2)}_\infty} &&=- L' \ket{\rho^{(1)}_\infty}.
 \label{mp2-2}
\end{eqnarray}

Before solving  Eqs.(\ref{mp2-1}, \ref{mp2-2}) for $\ket{\rho^{(1)}_\infty}$ and $\ket{\rho^{(2)}_\infty}$
let us   represent  the current~\eqref{current} in terms of nonequilibrium quasiparticles and understand which configurations in nonequilibrium density matrix give nonzero contribution to the expectation value of the current.
Using the transformation inverse to~\eqref{a_to_c}, we find
\begin{align}\label{current_c}
J_{\alpha} =& - i t \sum_{\sigma k n m} \bigr\{
 (\psi^{*}_{n, k \alpha} \varphi^*_{m} -\psi_n^*\varphi^*_{m,  k \alpha}  )\bra{I}  \widetilde c_{n \sigma} \widetilde c^{\dagger}_{m \sigma}\ket{\rho_\infty}
\nonumber
\\
& +i( \psi^*_{n, k \alpha} \psi_{m} -\psi^*_n \psi_{m, k\alpha}  )\bra{I} \widetilde c_{n \sigma} c_{m \sigma}\ket{\rho_\infty}\bigl\}.
\end{align}
The first term here is fully accounted within
the Hartree-Fock approximation
\begin{align}\label{current_HF_And1}
  J^{(0)}_\alpha &= -4t\mathrm{Im}\sum_{k n}\psi_{n,k\alpha}\varphi_n,
\end{align}
and does not contribute  to the expectation value over the correlated density matrix.
The $p$th-order perturbative theory correction to the HF current  comes from the second part of Eq.\eqref{current_c}
\begin{align}\label{current_p}
  J_\alpha^{(p)}
  =&2i  t\sum_{k mn} (\psi^*_{n,k\alpha}\psi_{m}-\psi^*_{n}\psi_{m,k\alpha}  )F^{(p)}_{mn}
  \notag\\
  =& -4t\mathrm{Im}\sum_{k mn}\psi^*_{n,k\alpha}\psi_{m}F^{(p)}_{mn},
\end{align}
where $F^{(p)}_{mn}$ is the expansion coefficient in
\begin{equation}
  \ket{\rho^{(p)}_\infty}=i\sum_{\sigma mn} F^{(p)}_{mn} c^\dag_{\sigma m}\widetilde c^\dag_{\sigma n}\ket{\rho^{(0)}_\infty}+\ldots
\end{equation}
and $F^{(p)}_{mn}=(F^{(p)}_{nm})^*$ as follows from  $\ket{\rho^{(p)}_\infty}=\widetilde{ \ket{\rho^{(p)}_\infty}}$.

Thus, to find the $p$th-order correction to the Hartree-Fock current we need to calculate $F^{(p)}_{mn}$.
Since $L'$ is normal ordered it does not
contain terms quadratic in nonequilibrium quasiparticle creation operators. Therefore, as it can be easily seen from Eq.(\ref{mp2-1}),   $F^{(1)}_{mn}=0$,
and the first nonvanishing correction to the current is $J_\alpha^{(2)}$.

 Eq.(\ref{mp2-1}) gives  the first-order perturbation theory correction to the Hartree-Fock steady state density matrix.
It contains the mixture of four nonequilibrium quasiparticle excitations:
\begin{equation}\label{rho_first}
  \ket{\rho^{(1)}_\infty}=\Bigl\{\sum_{klmn} G^{(1)}_{klmn}
 c^\dag_{\uparrow k}c^\dag_{\downarrow l}\widetilde  c^\dag_{\uparrow m}\widetilde c^\dag_{\downarrow n} \Bigr\} \ket{\rho^{(0)}_\infty},
\end{equation}
where
\begin{equation}
  G^{(1)}_{klmn}= - \frac{L^{(2)}_{klmn}}{\Omega_k+\Omega_l-\Omega^*_m -\Omega^*_n}
\end{equation}
and $G^{(1)}_{klmn}=G^{*(1)}_{mnkl}$.
Substituting~\eqref{rho_first} into (\ref{mp2-2})
we obtain
\begin{equation}\label{rho_second}
  \ket{\rho^{(2)}_\infty} = i\sum_{\sigma mn} F^{(2)}_{mn}
 c^\dag_{\sigma m}\widetilde  c^\dag_{\sigma n} \ket{\rho^{(0)}_\infty}+\ldots,
\end{equation}
where
\begin{equation}\label{F2}
  F^{(2)}_{mn} = \frac{1}{\Omega_m-\Omega^*_n}\sum_{ijk}\bigl\{ L^{(5)}_{mijk}G^{(1)}_{kjni} - L^{*(5)}_{nijk}G^{*(1)}_{kjmi}\bigr\}.
\end{equation}
Inserting~\eqref{F2} into~\eqref{current_p} we get the second-order perturbation theory correction to the Hartree-Fock current~\eqref{current_HF_And1}.

The omitted  terms in~\eqref{rho_second} contain four, six, and eight excited nonequilibrium quasiparticles
and these configurations do not contribute to the current in the second-order perturbation theory. However,  they contribute in higher orders.
We also note that each term in $\ket{\rho^{(p)}_\infty}$ contains  even number of excited quasiparticles and
the number of tilde and non-tilde creation
operators in the excited  quasiparticle configurations is the same  in all orders of perturbation theory.

\begin{figure*}[t!]
 \begin{centering}
\includegraphics[width=19cm]{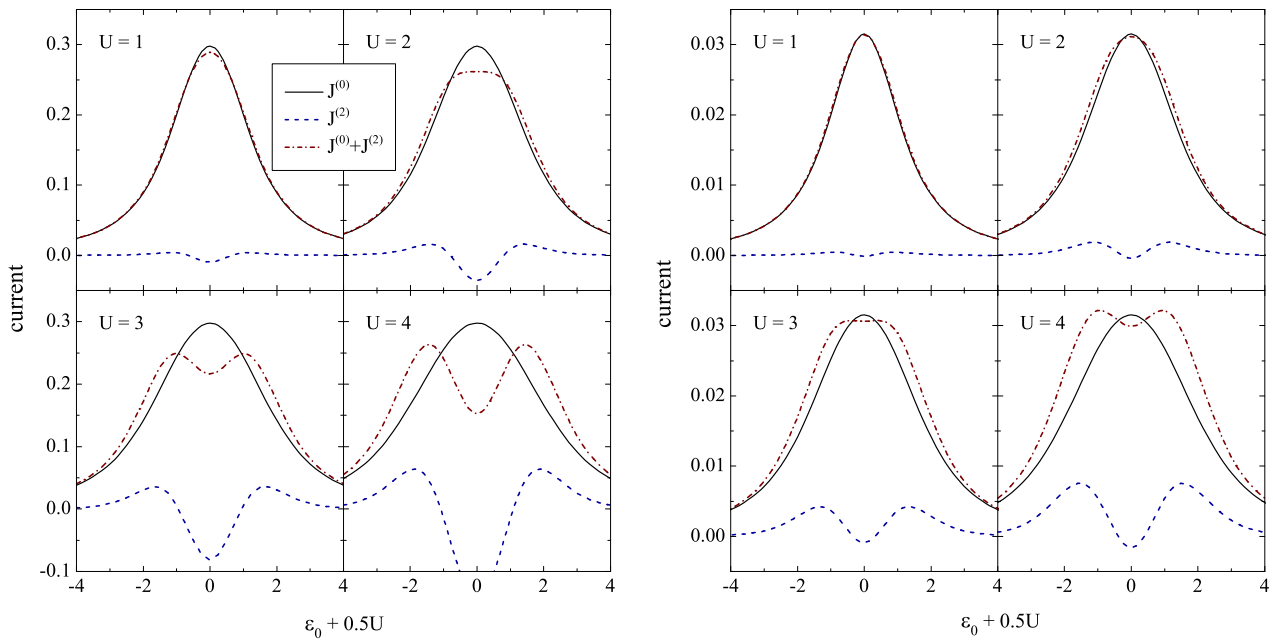}
\caption{(Color online):
 Current in the Anderson impurity model as a function of the central cite energy, $\varepsilon_0$, at  applied voltages  $V=1$ (left panels) and $0.1$ (right panels)
 for different  values of the Coulomb interaction energy $U$.}
\label{figure1}
 \end{centering}
\end{figure*}

To this point, the calculations are valid for arbitrary left and right leads. For the identical leads, Eqs.~(\ref{current_HF_And1},\ref{current_p}) can
be further simplified. Indeed, if we have identical left and right leads, i.e., if $E_{kL}=E_{kR}=E_k$, Eq.~\eqref{sys3} has $N$ eigenvalues given by $\Omega_n=E_l$.
For these eigenvalues $\psi_n=0$, and $\psi_{n,kL}=-\psi_{n,kR}=-\delta_{kl}2^{-1/2}$. The remaining $N+1$ eigenstates have  $\psi_{n,kL}=\psi_{n,kR}$ and hence they
do not contribute to the sum over $n$ in  Eqs.~(\ref{current_HF_And1},\ref{current_p}). Thus, we obtain
\begin{align}
\label{current_HF_And2}
  J^{(0)}_\alpha &= \mp2^{3/2}t\mathrm{Im}\sum_{n}\varphi_n,
  \\
 \label{current_p_symm}
 J^{(p)}_\alpha &=\mp2^{3/2}t\mathrm{Im}\sum_{mn}\psi_{m}F^{(p)}_{mn},
\end{align}
where the upper (lower) sign corresponds to $\alpha=L$ ($R$). In addition, because of $\psi_n=0$,  $L^{(5)}_{nijk}=0$ and the
second term in Eq.~\eqref{F2} vanishes.

Moreover, using explicit expression for $\varphi_n$ in the case of $\Omega_n=E_l$,
\begin{align}
\varphi_{n}= 2^{-1/2}
                  \cfrac{t(f_{lL}-f_{lR})}
         {(\varepsilon_{HF}-E_l)+2 t^2\sum\limits_{k}\cfrac{1}{(E_l-E^*_{k})}},
                     \end{align}
we can readily demonstrate that
if the
 macroscopically large part of the leads ($N\to\infty,~\gamma_k\to0$ )  is included into the embedded Hamiltonian~\eqref{embedded}, Eq.\eqref{current_HF_And2} becomes the Landauer formula
for the current in the Hartree-Fock approximation (see~\cite{dzhioev11a} for details).

\section{Numerical results}

We perform numerical calculations assuming that each lead has $N$ discrete, spin degenerate energy levels  uniformly distributed
in the bandwidth $[E_{min}:E_{max}]=[-5:5]$.  The tunneling
coupling strength $t$ is computed from the $\Gamma=2\pi\eta t^2=1$, where $\eta=N/(E_{max}-E_{min})$ is the density of states in the leads.
In all our calculations we take  $N=1600$ and $\gamma=2\Delta\varepsilon$, where $\Delta\varepsilon$
is the energy spacing between states in the leads.
Under this choice of the parameters  the imaginary part of the self-energy becomes energy independent constant $\Gamma=1$  within the bandwidth of the leads and vanishes outside the bandwidth. We use a symmetric applied voltage: $\mu_{L,R}=\pm 0.5V$.

To verify that the sufficient number of the leads levels is included into the embedded  Hamiltonian~\eqref{embedded}
we calculated the Hartree-Fock current for two different regimes, for fixed applied voltage  $V$ and varying
central cite energy $\varepsilon_0$ and vice versa, and then compared our results with  the  Landauer, i.e. exact,  mean-field current. We found the disagreement between obtained results is negligible (the maximum deviation  $< 1\%$ of the value of the current).

Fig.~\ref{figure1} shows the  current plotted as a function of the impurity level for different values of the Coulomb interaction energy $U$.
As we can see the HF current $J^{(0)}_L$ becomes broader with increasing of $U$ and reaches its maximum value at the symmetric point  $\varepsilon_0=-U/2$.
The second order perturbation theory correction to the current is
symmetric with respect to $\varepsilon_0 = - U/2$ and can be both positive and negative. We have numerically found that $J^{(2)}_L<0$ only when the Hartree-Fock level
$\varepsilon_{HF}$ is between $\mu_L$ and $\mu_R$.
For $U\lesssim 1$ the second order perturbation theory correction to the current is very small,  that means the Hartree-Fock approximation accounts for the main part of the electronic interactions.
For larger $U$  the second order perturbative corrections
change the behavior of the current both qualitatively and quantitatively. For example, when $U=3$ and $V=1$ the electronic correlations produce the dip in the current in the vicinity of the symmetric point $\varepsilon_0 = - U/2$.

We found that the larger the voltage (i.e. the further away we are from the equilibrium), the more important role the nonequilibrium electronic correlations play. This is evident from the comparison of the left (computed at $V=1$) and right (computed at $V=0.1$) panels of  Fig.~\ref{figure1}. It means that the nonequilibrium dynamics and electronic correlations are entangled in nontrivial way and  the nonequilibrium enhances and amplifies the role of electronic correlations.

\section{Conclusions}

In this paper, we developed the second-order post-Hartree-Fock perturbation theory for the electron current through a  region of out of equilibrium, interacting electrons. As an example, we considered electron transport through out of equilibrium Anderson model, although all our derivations are also directly applied to a general
molecular junctions described by full many-electron Hamiltonian. We started with the Lindblad kinetic equation for the embedded molecular system. We converted the kinetic equation to super-fermion representation and define nonequilibrium quasiparticles within Hartree-Fock approximation. Then we applied the Wick theorem and perform the normal ordering of the Liouvillian with respect to the vacuum for nonequilibrium quasiparticles. We developed the second-order post-Hartree-Fock perturbation theory by admixing two- and four-quasiparticle configurations to the nonequilibrium vacuum. Our numerical calculations  for out of  equilibrium Anderson impurity demonstrated that nonequilibrium electronic correlations  may produce significant quantitative and qualitative corrections to  Hartree-Fock electronic transport properties. We also found that  the nonequilibrium enhances the role of electronic correlations.

\begin{acknowledgments}
 This work has been supported by the Francqui Foundation, Belgian Federal Government under the Inter-university Attraction Pole project NOSY  and
 Programme d'Actions de Recherche Concert\'ee de la Communaut\'e francaise (Belgium).

\end{acknowledgments}

\vspace{2ex}
\appendix

\section{Normal ordered Liouvillian in the basis of nonequilibrium Hartree-Fock quasiparticles}
Here we give the explicit expression of $L'$ in terms of nonequilibrium quasiparticle creation and annihilation operators.
Using the inverse transformation
\begin{align}
  a^\dag_\sigma&=\sum_n (\psi_{n}-\varphi_{n})c^\dag_{\sigma n} + i\sum_n \psi^*_{n}\widetilde c_{\sigma n},
  \notag\\
  a_\sigma&=\sum_n \psi_{n} c_{\sigma n} - i \sum_n \varphi^*_{n} \widetilde c^\dag_{\sigma n}
\end{align}
we obtain
\begin{align}
L'=&U:(a^\dag_\uparrow a_\uparrow a^\dag_\downarrow a_\downarrow - \mathrm{t.c.}):
&\nonumber \\
=&\sum\limits_{klmn}\Bigl\{ (L^{(1)}_{klmn}c^\dag_{k_\uparrow} c^\dag_{l_\downarrow} c_{m_\downarrow} c_{n_\uparrow} - \mathrm{t.c.})
&\nonumber \\
&+ L^{(2)}_{k l m n} c^\dag_{k\uparrow} c^\dag_{l_\downarrow} \widetilde c^\dag_{m_\uparrow} \widetilde c^\dag_{n_\downarrow}
 \nonumber \\
&+ L^{(3)}_{klmn}(c^\dag_{k_\uparrow}\widetilde c^\dag_{l_\downarrow} \widetilde c_{m_\downarrow}  c_{n_\uparrow}+
                  c^\dag_{k_\downarrow}\widetilde c^\dag_{l\uparrow} \widetilde c_{m_\uparrow}  c_{n_\downarrow}
&\nonumber \\
&~~~-c^\dag_{k_\uparrow}\widetilde c^\dag_{l_\uparrow} \widetilde c_{m_\downarrow}  c_{n_\downarrow}-
  c^\dag_{k_\downarrow}\widetilde c^\dag_{l_\downarrow} \widetilde c_{m_\uparrow}  c_{n_\uparrow})
\nonumber  \\
&+i\bigl[L^{(4)}_{klmn}(c^\dag_{k_\uparrow} c^\dag_{l_\downarrow}\widetilde c^\dag_{m_\downarrow} c_{n_\uparrow}+
  c^\dag_{k_\downarrow} c^\dag_{l_\uparrow}\widetilde c^\dag_{m_\uparrow} c_{n_\downarrow}) + \mathrm{t.c.}\bigr]
 \nonumber \\
&+i\bigl[(L^{(5)}_{klmn}c^\dag_{k_\uparrow} \widetilde c_{l_\downarrow} c_{m_\downarrow} c_{n_\uparrow}+
  +c^\dag_{k_\downarrow}\widetilde c_{l_\uparrow} c_{m_\uparrow} c_{n_\downarrow}+\mathrm{t.c.}\bigr]\Bigr\},
  \end{align}
where $L^{(i)}_{klmn}$ are given by
\begin{align*}
  L^{(1)}_{klmn}=& U(\psi_k\psi_l-\psi_k\varphi_l-\varphi_k\psi_l)\psi_m\psi_n,
   \\
  L^{(2)}_{klmn}=&U\bigl[(\psi_k-\varphi_k)(\psi_l-\varphi_l)\varphi^*_m\varphi^*_n
   \\
  &-\varphi_k\varphi_l(\psi^*_m-\varphi^*_m)(\psi^*_n-\varphi^*_n)\bigr],
  \\
  L^{(3)}_{klmn}=&U(\varphi_k\psi^*_l-\psi_k\varphi^*_l)\psi^*_m\psi_n,
   \\
  L^{(4)}_{klmn}=&-U\bigl[(\psi_k-\varphi_k)(\psi_l-\varphi_l)\varphi^*_m
   \\
&+ \varphi_k\varphi_l(\psi^*_m-\varphi^*_m)\bigr]\psi_n,
   \\
  L^{(5)}_{klmn}=&U\psi_k\psi_l^*\psi_m\psi_n.
\end{align*}
It is notable that because of $\bra{I}L=0$, $L'$ does not contain terms of four annihilation operators.


\begin{thebibliography}{13}
\expandafter\ifx\csname natexlab\endcsname\relax\def\natexlab#1{#1}\fi
\expandafter\ifx\csname bibnamefont\endcsname\relax
  \def\bibnamefont#1{#1}\fi
\expandafter\ifx\csname bibfnamefont\endcsname\relax
  \def\bibfnamefont#1{#1}\fi
\expandafter\ifx\csname citenamefont\endcsname\relax
  \def\citenamefont#1{#1}\fi
\expandafter\ifx\csname url\endcsname\relax
  \def\url#1{\texttt{#1}}\fi
\expandafter\ifx\csname urlprefix\endcsname\relax\def\urlprefix{URL }\fi
\providecommand{\bibinfo}[2]{#2}
\providecommand{\eprint}[2][]{\url{#2}}

\bibitem[{\citenamefont{Jensen}(2006)}]{jensen-book}
\bibinfo{author}{\bibfnamefont{F.}~\bibnamefont{Jensen}},
  \emph{\bibinfo{title}{Introduction to Computational Chemistry}}
  (\bibinfo{publisher}{Willey}, \bibinfo{year}{2006}).

\bibitem[{\citenamefont{Thygesen and Rubio}(2008)}]{PhysRevB.77.115333}
\bibinfo{author}{\bibfnamefont{K.~S.} \bibnamefont{Thygesen}} \bibnamefont{and}
  \bibinfo{author}{\bibfnamefont{A.}~\bibnamefont{Rubio}},
  \bibinfo{journal}{Phys. Rev. B} \textbf{\bibinfo{volume}{77}},
  \bibinfo{pages}{115333} (\bibinfo{year}{2008}).

\bibitem[{\citenamefont{My\"oh\"anen et~al.}(2009)\citenamefont{My\"oh\"anen,
  Stan, Stefanucci, and van Leeuwen}}]{PhysRevB.80.115107}
\bibinfo{author}{\bibfnamefont{P.}~\bibnamefont{My\"oh\"anen}},
  \bibinfo{author}{\bibfnamefont{A.}~\bibnamefont{Stan}},
  \bibinfo{author}{\bibfnamefont{G.}~\bibnamefont{Stefanucci}},
  \bibnamefont{and} \bibinfo{author}{\bibfnamefont{R.}~\bibnamefont{van
  Leeuwen}}, \bibinfo{journal}{Phys. Rev. B} \textbf{\bibinfo{volume}{80}},
  \bibinfo{pages}{115107} (\bibinfo{year}{2009}).

\bibitem[{\citenamefont{Spataru et~al.}(2009)\citenamefont{Spataru, Hybertsen,
  Louie, and Millis}}]{PhysRevB.79.155110}
\bibinfo{author}{\bibfnamefont{C.~D.} \bibnamefont{Spataru}},
  \bibinfo{author}{\bibfnamefont{M.~S.} \bibnamefont{Hybertsen}},
  \bibinfo{author}{\bibfnamefont{S.~G.} \bibnamefont{Louie}}, \bibnamefont{and}
  \bibinfo{author}{\bibfnamefont{A.~J.} \bibnamefont{Millis}},
  \bibinfo{journal}{Phys. Rev. B} \textbf{\bibinfo{volume}{79}},
  \bibinfo{pages}{155110} (\bibinfo{year}{2009}).

\bibitem[{\citenamefont{Dahnovsky}(2009)}]{PhysRevB.80.165305}
\bibinfo{author}{\bibfnamefont{Y.}~\bibnamefont{Dahnovsky}},
  \bibinfo{journal}{Phys. Rev. B} \textbf{\bibinfo{volume}{80}},
  \bibinfo{pages}{165305} (\bibinfo{year}{2009}).

\bibitem[{\citenamefont{Thygesen and Rubio}(2007)}]{thygesen07}
\bibinfo{author}{\bibfnamefont{K.~S.} \bibnamefont{Thygesen}} \bibnamefont{and}
  \bibinfo{author}{\bibfnamefont{A.}~\bibnamefont{Rubio}}, \bibinfo{journal}{J.
  Chem. Phys.} \textbf{\bibinfo{volume}{126}}, \bibinfo{pages}{091101}
  (\bibinfo{year}{2007}).

\bibitem[{\citenamefont{Darancet et~al.}(2007)\citenamefont{Darancet, Ferretti,
  Mayou, and Olevano}}]{PhysRevB.75.075102}
\bibinfo{author}{\bibfnamefont{P.}~\bibnamefont{Darancet}},
  \bibinfo{author}{\bibfnamefont{A.}~\bibnamefont{Ferretti}},
  \bibinfo{author}{\bibfnamefont{D.}~\bibnamefont{Mayou}}, \bibnamefont{and}
  \bibinfo{author}{\bibfnamefont{V.}~\bibnamefont{Olevano}},
  \bibinfo{journal}{Phys. Rev. B} \textbf{\bibinfo{volume}{75}},
  \bibinfo{pages}{075102} (\bibinfo{year}{2007}).

\bibitem[{\citenamefont{Gurvitz and Prager}(1996)}]{gurvitz96}
\bibinfo{author}{\bibfnamefont{S.~A.} \bibnamefont{Gurvitz}} \bibnamefont{and}
  \bibinfo{author}{\bibfnamefont{Y.~S.} \bibnamefont{Prager}},
  \bibinfo{journal}{Phys. Rev. B} \textbf{\bibinfo{volume}{53}},
  \bibinfo{pages}{15932} (\bibinfo{year}{1996}).

\bibitem[{\citenamefont{Harbola et~al.}(2006)\citenamefont{Harbola, Esposito,
  and Mukamel}}]{PhysRevB.74.235309}
\bibinfo{author}{\bibfnamefont{U.}~\bibnamefont{Harbola}},
  \bibinfo{author}{\bibfnamefont{M.}~\bibnamefont{Esposito}}, \bibnamefont{and}
  \bibinfo{author}{\bibfnamefont{S.}~\bibnamefont{Mukamel}},
  \bibinfo{journal}{Phys. Rev. B} \textbf{\bibinfo{volume}{74}},
  \bibinfo{pages}{235309} (\bibinfo{year}{2006}).

\bibitem[{\citenamefont{Schmutz}(1978)}]{schmutz78}
\bibinfo{author}{\bibfnamefont{M.}~\bibnamefont{Schmutz}}, \bibinfo{journal}{Z.
  Physik. B} \textbf{\bibinfo{volume}{30}}, \bibinfo{pages}{97 }
  (\bibinfo{year}{1978}).

\bibitem[{\citenamefont{Prosen}(2008)}]{prosen08}
\bibinfo{author}{\bibfnamefont{T.}~\bibnamefont{Prosen}}, \bibinfo{journal}{New
  Journal of Physics} \textbf{\bibinfo{volume}{10}}, \bibinfo{pages}{043026}
  (\bibinfo{year}{2008}).

\bibitem[{\citenamefont{Harbola and Mukamel}(2008)}]{Harbola2008}
\bibinfo{author}{\bibfnamefont{U.}~\bibnamefont{Harbola}} \bibnamefont{and}
  \bibinfo{author}{\bibfnamefont{S.}~\bibnamefont{Mukamel}},
  \bibinfo{journal}{Physics Reports} \textbf{\bibinfo{volume}{465}},
  \bibinfo{pages}{191 } (\bibinfo{year}{2008}).

\bibitem[{\citenamefont{Dzhioev and Kosov}(2011)}]{dzhioev11a}
\bibinfo{author}{\bibfnamefont{A.~A.} \bibnamefont{Dzhioev}} \bibnamefont{and}
  \bibinfo{author}{\bibfnamefont{D.~S.} \bibnamefont{Kosov}},
  \bibinfo{journal}{J. Chem. Phys.} \textbf{\bibinfo{volume}{134}},
  \bibinfo{pages}{044121} (\bibinfo{year}{2011}).

\end{thebibliography}
\end{document}